\documentclass[aps,prb,twocolumn,groupedaddress,longbibliography]{revtex4-2}
\usepackage{graphicx}
\usepackage{hyperref}
\usepackage{dcolumn}
\usepackage{bm}
\usepackage{romannum}

\begin{document}


\title{Normal-state resistivity and the depairing current density of BaFe$_2$(As,P)$_2$ nanobridges along the $c$ axis}


\author{Yuki Mizukoshi$^1$, Kotaro Jimbo$^1$, Akiyoshi Park$^2$, Yue Sun$^3$, Tsuyoshi Tamegai$^2$, Haruhisa Kitano$^1$}
\affiliation{
$^1$Department of Physics, Aoyama Gakuin University, Sagamihara 252-5258, Japan\\ $^2$Department of Applied Physics, The University of Tokyo, Tokyo 113-8656, Japan\\$^3$Department of Physics, Southeast University, Nanjing 211189, China 
}


\date{\today}

\begin{abstract}
We report precise measurements to obtain the normal-state resistivity and the depairing current density of BaFe$_2$(As$_{1-x}$P$_x$)$_2$($x\sim0.29-0.32$) nanobridges along the $c$ axis, 
by using focused ion beam (FIB) techniques. 
We obtained both of the $ab$-plane and $c$-axis resistivity ($\rho_{ab}$ and $\rho_c$) in the same part of a specimen, by fabricating the $c$-axis nanobridge in the middle of a narrow bridge extended in $ab$-plane, in spite of the slight deficiency of P dopant due to the additional FIB fabrication. 
The normal-state resistivity anisotropy agreed with the previous results for bulk samples,  showing $\rho_c/\rho_{ab} < 8$ just above the superconducting transition temperature, $T_c$, in the slightly underdoped region and a slight decrease with increasing temperatures. 
The critical current density obtained in the $c$-axis nanobridges near the optimal doping reaches $\sim$8~MA/cm$^2$ at 0.15$T_c$, corresponding to about 87\% of a depairing limit derived by the Eilenberger equations. 
An extrapolation to $T=$0~K using the Ginzburg-Landau model suggests that 
the anisotropy of the depairing current density roughly corresponds to that of the normal-state resistivity. 
At low temperatures, we also observed a step-like voltage jump before arriving at the depairing limit, suggesting the occurrence of phase-slip phenomena near the depairing processes. 

\end{abstract}


\maketitle

\section{Introduction}
The phase diagram of BaFe$_2$As$_2$-based superconductors has been extensively investigated to explore key fluctuations giving rise to high temperature superconductivity in iron-based superconductors (IBSs) \cite{Fernandes2014}. 
There are three phase diagrams in K, Co, and P-doped BaFe$_2$As$_2$, which are regarded as the hole-doped, electron-doped, and isovalent substituted system, respectively. 
In these systems, the superconducting (SC) transition temperature, $T_c$, has a dome-shaped dependence on the concentration of K, Co and P and the SC phase is in proximity to the antiferromagnetic ordered phase, showing strong similarity to cuprate superconductors. 

However, in contrast to cuprates, the relative low anisotropy of transport properties in the normal-state and superconducting state was reported by several studies on the electron-doped BaFe$_2$As$_2$ \cite{Tanatar2009,Tanatar2009b,Tanatar2011} and P-doped BaFe$_2$As$_2$ \cite{Tanatar2013}. 
In addition, the softness of this material makes the determination of transport properties along the $c$ axis quite difficult, since partial cracks introduced by cutting and exfoliating single crystals influence the current distribution seriously \cite{Tanatar2009}. 
Thus, the normalized data of the $c$-axis resistivity by the measured value at 300~K were often plotted in the previous studies \cite{Tanatar2011,Tanatar2013}.  

Recently, we demonstrated that a narrow bridge structure fabricated by using focused ion beam (FIB) techniques was quite useful to determine the normal-state resistivity anisotropy of Fe$_{1+y}$Te$_{1-x}$Se$_x$ single crystals \cite{Sun2021}. 
In contrast to the conventional methods using two samples with different electrode for the resistivity measurements in the $ab$-plane and along the $c$-axis ($\rho_{ab}$ and $\rho_c$, respectively), the advantage of FIB method is that both of $\rho_{ab}$ and $\rho_c$ can be measured in the same region of a specimen, by additionally making a short nanobridge along the $c$ axis into the narrow bridge extended to the $ab$-plane. 
In the previous study on Fe$_{1+y}$Te$_{0.6}$Se$_{0.4}$\cite{Sun2021}, we found that the normal-state anisotropy determined by the ratio of $\rho_c/\rho_{ab}$ was systematically decreased with the removal of interstitial excess iron. 

By using the same narrow bridge structure, we also succeeded in measuring the critical current density, $j_c$, for the depairing of Cooper pairs moving along the $c$ axis of Fe$_{1+y}$Te$_{1-x}$Se$_x$ single crystals \cite{Sun2020}. 
The transport $j_c$ for the $c$-axis microbridge with a width smaller than the Pearl length reached 1.3~MA/cm$^2$ at $\sim 0.3T_c$. This value is at least one order of magnitude larger than the depinning $j_c$, obtained by magnetization hysteresis loop (MHL) measurements, 
while the result for transport $j_c$ is comparable to a calculated value of the depairing $j_c$ based on the Ginzburg-Landau (GL) theory. 
The temperature dependence of $j_c$ showed a quantitative agreement with the GL theory down to 0.83$T_c$ and a qualitative agreement with Kupriyanov-Lukichev (KL) theory \cite{KLtheory} down to $\sim 0.3T_c$. In the KL theory, the contribution of a growth of the SC gap is considered through a numerical solution of Eilenberger equations. 

In this study,  we applied the FIB method to BaFe$_2$(As$_{1-x}$P$_x$)$_2$ ($x=0.32$) single crystals, in order to measure both $\rho_{ab}$ and $\rho_c$ as well as the depairing $j_c$ along the $c$ axis. 
We found that this method facilitated the detection and avoidance of partial cracks affecting the local current distribution, compared to the conventional method using bulk samples.  
The magnitude and temperature dependence of $\rho_{ab}$ show an excellent agreement with the previously-reported results using bulk crystals \cite{Kasahara2010}. 
On the other hand, $\rho_c(T)$ shows a slightly negative curvature in $\rho_c(T)$ at around 250~K, indicating that the concentration of P after the additional FIB fabrication is slightly in the underdoped side of the SC dome \cite{Tanatar2013}. 
We actually confirmed the slight decrease of $x$ (less than 0.01) using the energy-dispersive X-ray (EDX) analyses for the fabricated narrow bridges. 
The normal-state anisotropy $\rho_c/\rho_{ab}$ for the slightly underdoped side near the optimal doping ($x\sim$0.32) is at most 8 just above $T_c$ and shows a slight decrease with increasing temperature. The magnitude of $\rho_c/\rho_{ab}$ semiquantitatively agrees with the previously-reported results for bulk crystals \cite{Tanatar2013}.

We also found that the $c$-axis transport $j_c$ for the nanobridge showing the highest $T_c$(=28~K) reached 8~MA/cm$^2$ at $\sim$0.15$T_c$. It is at least 9 times larger than the depinning $j_c$ in the $ab$-plane determined by MHL for the optimally-doped bulk crystals with $x=0.33$ \cite{Ishida2017,Park2020}, and corresponds to $\sim$~87~\% of the depairing $j_c$ derived from the KL theory. 
At low temperatures, we also discovered a sharp jump in the current-voltage ($I$-$V$) characteristics just below the depairing $j_c$, suggesting the contribution of phase slip phenomena near the depairing limit.

\section{Experiment}
Single crystals of BaFe$_2$(As$_{1-x}$P$_x$)$_2$ were grown by using Ba$_2$As$_3$/Ba$_2$P$_3$ flux \cite{Park2020}. 
Several samples with $x=0.32$ are carefully cleaved and cut into rectangular shape (typically, 700~$\mu$m$\times$150~$\mu$m$\times$10-20~$\mu$m). 
The value of $x$ for the cleaved samples before the FIB microfabrication was determined by energy-dispersive X-ray (EDX) analysis and the $c$-axis lattice parameter determined by the (00$l$) peaks in the X-ray diffraction (XRD) pattern \cite{Park2020}. 
The cleaved sample is glued on a sapphire substrate using an epoxy (Stycast 1266). 
Four Au electrodes are prepared on each sample by using a sputtering technique. 

A narrow bridge is fabricated between two voltage electrodes by using the FIB etchings (Hitachi High-Tech MI4050), as shown in Fig.~1(a). 
The existence of partial cracks is directly checked by observing scanning ion microscope (SIM) images of the bridge. 
Even if a tiny crack hides into the narrow bridge part, we can easily find its influence through the dc resistance measurements down to low temperatures at a rate of 15-20~K/h, since the narrow bridge including such tiny cracks cannot suffer stress due to thermal shrinkage. 
After measuring the dc resistance ($R_1$) for the microbridge along the $ab$-plane, we fabricate the $c$-axis nanobridge by adding two slits, as shown in Fig.~1(b). 
Then, we measure the dc resistance ($R_2$) for the $c$-axis nanobridge. 
By using $R_1$ and $R_2$, both of $\rho_{ab}$ and $\rho_c$ in the same region of the microbridge can be obtained, as described in the previous study \cite{Sun2021}. 
As described in Supplementary Information \cite{SeeSI,Ota2009,Kakizaki2017}, we went through the limitation of this method for materials with small anisotropy. 
Finally, the transport $j_c$ is measured for the $c$-axis nanobridge by using a combination of pulse current source and nanovoltmeter (Delta mode of Keithley 6221/2182A) \cite{Sun2020}. The pulse width and the repetition period are 100~$\mu$s and 2$\sim$3~s, respectively. 

In this paper, we focus on 5 nanobridges, as listed in Table I. 
Here, the value of $x$ for each bridge was determined by EDX analyses (Zeiss ULTRA55/Bruker QUANTAX) after measuring the transport $j_c$. 
The value of $T_c$ was determined by a maximum temperature of the zero-resistance in $R_2$ measurements ($T_c^R$) or the onset temperature of the critical current density ($J_c$) in the $J_c$-$V$ characteristics ($T_c^J$). 
We found that the FIB microfabrication removed P dopants slightly in the bridge part of BaFe$_2$(As$_{1-x}$P$_x$)$_2$ single crystals, leading to a distribution of $T_c$ in the fabricated bridge samples, as shown in Table I \cite{Note1}.   

\begin{table*}
\caption{\label{table1}Sizes of the fabricated bridges and properties in the SC state. $x$ is the concentration of P, determined by EDX analyses for three points on the side wall of the narrow bridges. $w$ and $h$ are the lateral sizes of the nanobridge along the $c$ axis, and $l$ is a length of the nanobridge, as shown in Fig.~1. $T_c^R$ and $T_c^J$ are determined by a rising temperature of $R_2(T)$ and $j_c(T)$, respectively. $j_c^{\rm GL}(0)$ is an extrapolated value by the fitting of the measured data just below $T_c$ to a behavior of $(1-T/T_c)^{1.5}$ in the GL theory. }
\begin{ruledtabular}
\begin{tabular}{c c c c c c c c} 
sample & $x$ & $w~(\mu$m) & $h~(\mu$m) & $l~(\mu$m) & $T_c^R$ (K) & $T_c^J$ (K) & $j_c^{\rm GL}(0)$ (MA/cm$^2$) \\
bp06 & 0.324 & 0.3 & 1.3 & 0.8 & 27 & 27 & 17.6 \\
bp07 & 0.317 & 0.9 & 0.8 & 0.4 & 26 & 26 & 6.0 \\
bp16 & 0.287  & 0.3 & 1.4 & 2.0 & 16 & 18 & 14.7 \\
bp18 & 0.320 & 0.3 & 1.0 & 0.2 & 26 & 27.5 & 15.2 \\
ij12 & 0.321$\pm$0.001 & 0.6 & 0.8 & 0.2 & 28 & 28 & 24.3 \\
 \end{tabular}
 \end{ruledtabular}
 \end{table*}

\section{Results and Discussion} 
Figures 1(a) and 1(b) show the temperature dependence of $R_1$ and $R_2$ for a sample (bp07), respectively. 
As described in the previous study \cite{Sun2021}, the value of $R_1$ is given by a sum of several resistances in the $ab$ plane ($R_{ab}$), while there is a resistance along the $c$ axis ($R_c$) as well as the sum of $R_{ab}$ in the measurements of $R_2$. 
Note that the contribution of $R_c$ is enhanced by the small cross-sectional area of the $c$-axis nanobridge, compared to the conventional Corbino electrode configuration to obtain the out-of-plane resistivity in a plate-like sample. 
Thus, $\rho_{ab}$ and $\rho_c$ are obtained with sufficient accuracy from $R_1$ and $R_2$, as shown in Fig.~2(a). 
The magnitude and the temperature dependence of $\rho_{ab}$ show almost perfect agreement with the previous results for the optimally-doped sample ($x$=0.33) \cite{Kasahara2010}. 
On the other hand, the temperature dependence of $\rho_c$ seems to be slightly underdoped region in the SC dome, since a small negative curvature is observed in $\rho_c(T)$ at around 250~K. 
In the previous study using bulk crystals \cite{Tanatar2013}, only a $T$-linear behavior was observed in $\rho_c(T)$ with $x$=0.33, while $\rho_c(T)$ in the underdoped samples showed a shallow maximum at higher temperatures. 
In addition, it is found that the superconducting transition observed in $\rho_c$ occurs at a slightly lower temperature than that observed in $\rho_{ab}$. 
EDX analyses for this sample also indicate a slight decrease of $x$ ($<1$\%). 
These results suggest an influence of the slight deficiency of P dopant due to the additional FIB fabrication. 

The normal-state resistivity anisotropy is at most 8 just above $T_c$, as shown in Fig.~2(b). 
Note that this value is presumably overestimated, since the slight decrease of $x$ is expected to push up $\rho_c(T)$. 
Nevertheless, the obtained anisotropy of $\rho_c/\rho_{ab}$ seems to agree with the previous results for a bulk crystal of $x$=0.33 ($\rho_c/\rho_{ab}\approx6\pm2$)\cite{Tanatar2013}.  
We also investigated the dependence of $\rho_c$ on the length of nanobridge along the $c$ axis \cite{SeeSI}, and found that the additional FIB fabrication to lengthen the nanobridge leads to an increase $R_{ab}$ as well as $R_c$, since the elongation of a slit made in the side wall of narrow bridge gives rise to both an increase of the length of the $c$-axis nanobridge and a decrease of the thickness of in-plane bridge. 
Thus, in the case of small anisotropy of $\rho_c/\rho_{ab}$, the nanobridge with a shorter length is more useful to obtain the contribution of $R_c$. 
 
Figure 3(a) shows the $I$-$V$ characteristics of another sample (ij12) measured at several temperatures below $T_c$, which shows the maximal values of $T_c$ and $j_c$ among the measured samples. 
$R_2(T)$ measurements for this sample showed almost $T$ linear behavior and no shallow maximum at higher temperatures, suggesting that this sample (ij12) was actually near the optimal doping. 
In contrast to other samples, we observed two-staged voltage jump at low temperatures below 14~K, as shown in Fig.~3(a). 
After the first $I$-$V$ measurements shown in Fig.~3(a), we performed the 2nd measurements below 14~K again, in order to check the reproducibility of the two-staged voltage jump.  
As shown in Fig.~3(b), we found the stochastic nature of the 1st voltage jump observed at a lower current, since the current position corresponding to the 1st jump was decreased for $T$=10~K and 8~K, while it was increased for $T$=6~K. 
On the other hand, we found that the variation of the current positions corresponding to the 2nd jump between two measurements was much smaller than those for the 1st jump. 
Thus, we determined the critical current, $I_c$, by using two criteria of 10~$\mu$V and 50~$\mu$V, as shown in Fig.~3(c). 
Here, we used the firstly-measured data shown in Fig.~3(a), since the SC properties in the 2nd $I$-$V$ measurements seemed to be slightly degraded. 
The magnitude of $j_c$ was obtained by using a cross-sectional area of the $c$-axis nanobridges, given by a product of $w\times h$. 
Since there wa no report on $\lambda_L$ along the $c$ axis for the optimally doped BaFe$_2$(As$_{1-x}$P$_x$)$_2$, we roughly estimated the Pearl length ($\Lambda=2\lambda_L^2/h$) by assuming that $\lambda_L\sim$1~$\mu$m. 
Note that this value is comparable to the reported value for $\lambda_L$ along the $c$ axis in the optimally doped Ba(Fe$_{1-x}$Co$_x$)$_2$As$_2$ \cite{Prozorov2011}. 
By using such an estimation, we confirmed that the cross-sectional area ($\sim$0.48~$\mu$m$^2$ for sample ij12) of the $c$-axis nanobridge was smaller than $2\lambda_L^2$ along the $c$ axis. Thus, it is considered that the spatial distribution of the supercurrent flowing along the $c$-axis in the fabricated nanobridges is almost homogeneous \cite{Sun2020,Nawaz2013}, except for sample bp07 with a larger cross-sectional area than that of sample ij12. 

As shown in Fig.~3(c), the temperature dependence of $j_c$ determined by the lower criterion of 10~$\mu$V shows a small dip between 10~Kand 15~K, which is probably due to the two-staged jumps occurring below 14~K. 
On the other hand, $j_c(T)$ determined by the higher criterion of 50~$\mu$V shows a monotonic increase down to the lowest temperature and seems to be more smoothly connected to $j_c(T)$ above 15~K, which was obtained by the threshold of 10~$\mu$V. 
This suggests that the second jumps observed below 14~K are closer to the depairing limit and that the first jumps are rather caused by another origin, as discussed below. 
By adopting the values of $j_c$ at the second jumps below 12~K, we compared the temperature dependence of $j_c$ normalized by $j_c(0)$ with the GL and KL theories, as shown in Fig.~3(d). Here, $j_c^{\rm GL}(0)$(=24.3~MA/cm$^2$) was determined by an extrapolation of the fitting data obtained just below $T_c$ to a behavior of $(1-T/T_c)^{1.5}$ predicted by the GL theory. 
We compared this value with a calculated value of $j_c^{\rm GL}(0)$(=79 MA/cm$^2$\cite{Sun2020}) in the $ab$ plane, based on the GL theory ($j_c=c\phi_0/12\sqrt{3}\pi^2\xi\lambda^2$). 
The obtained anisotropy ($j_c^{ab}(0)/j_c^c(0)\sim$3.3) in the SC state seems to be slightly larger than the normal-state anisotropy determined by $\sqrt{\rho_c/\rho_{ab}}$. 
Note that the depairing current density is proportional to a critical velocity of Cooper pairs transporting the SC current, which is determined by a crossover of the kinetic energy of Cooper pairs with the SC condensation energy \cite{Clem2012}. Thus, the anisotropy of $j_c^{ab}/j_c^c$ is expected to be equal to a square root of the anisotropy of the effective mass, $\sqrt{m_c/m_{ab}}$. 
However, since IBS is a multiband system with several Fermi surfaces, an issue of the anisotropy in the charge transport properties should be more carefully discussed. 

Next, we discuss $j_c(T)/j_c^{\rm GL}(0)$ by comparing with the GL and KL theories. 
As shown in Fig.~3(d), $j_c(T)/j_c(0)$ quantitatively agrees with both theories down to $\sim$0.8~$T_c$ and with the KL theory down to $\sim$0.55~$T_c$. 
The experimental result of $j_c$ obtained at the lowest temperature ($T\sim$0.15~$T_c$) is $\sim$8~MA/cm$^2$, which is about 87\% of a value obtained by the KL theory and about an order of magnitude larger than the depinning $j_c$ determined by MHL \cite{Ishida2017,Park2020}. 
These results strongly suggest that the obtained $j_c(T)$ actually represents the depairing current density along the $c$ axis of BaFe$_2$(As$_{1-x}$P$_x$)$_2$ near the optimal doping and that the depairing current density for single-crystalline samples was experimentally explored down to $\sim$0.15~$T_c$.
This temperature range is larger than that for the $c$-axis depairng current density of Fe$_{1+y}$Te$_{0.6}$Se$_{0.4}$ single crystals (down to $\sim$0.3~$T_c$) \cite{Sun2020} and that for the $ab$-plane depairing current density of the optimally doped (Ba$_{0.5}$K$_{0.5}$)Fe$_2$As$_2$ single-crystalline microbridges (down to $\sim$0.87~$T_c$) \cite{Li2013}.  
In addition, the temperature dependence of $j_c(T)/j_c^{\rm GL}(0)$ showing a better agreement with the KL theory indicates that the first jump in the two-staged voltage jumps observed at low temperatures is rather caused by a different origin from the depairing events. 

We consider that the first jump is attributed to phase-slip (PS) phenomenon, which has been often observed in narrow SC wires \cite{Skocpol1974,Tidecks1990}, SC strips \cite{Sivakov2003} and the two-dimensional superconductors with a few layers \cite{Paradiso2019}. 
The characteristic step structure in the $I$-$V$ curves was observed in the SC filaments just below $T_c$ and has been successfully explained by the formation of phase-slip centers (PSCs) and the interactions between PSCs \cite{Skocpol1974,Tidecks1990}. 
The two-dimensional analogue of PSC has been observed in wide strips and films \cite{Sivakov2003,Paradiso2019} and is called as phase-slip line (PSL), characterized by the formation of kinematic vortices or kinematic vortex-antivortex pairs  \cite{Andronov1993,Berdiyorov2009}.  

Unfortunately, the details of such resistive SC states existing between the SC state with zero voltage and the fully normal-state are still poorly understood. 
In numerical studies based on the time-dependent GL (TDGL) theory and its generalization \cite{Kramer1977,Kramer1978}, oscillatory PS solutions intrinsically emerge both below and above the depairing current and the system exhibits a hysteretic behavior. 
This suggests an intrinsic instability of the SC filament transporting the SC current with a high speed. 
In this work,  the two-staged steps were observed at low temperatures, in contrast to early studies on narrow SC wires \cite{Skocpol1974,Tidecks1990}, where the consecutive steps appeared just below $T_c$. 
In addition, the presence or absence of the step structure was dependent on samples, implying the stochastic and unstable nature of the resistive SC state near the depairing limit. 
These results clearly show that the PS phenomena with stochastic and unstable nature occur even in the three-dimensional nanobridges made of IBS single crystals. We will discuss this issue in more depth in the near future. 

We compared the temperature dependence of $j_c$ between the fabricated samples except for the sample (bp07), which had the largest width ($w$=0.9~$\mu$m) and was used to obtain the normal-state anisotropy. 
The value of $j_c^{\rm GL}(0)$ for bp07 was much smaller than other samples, as shown in Table I.  
Although the magnitude of $j_c(<$2~MA/cm$^2$) at the lowest temperature was still larger than the in-plane depinning $j_c$ \cite{Park2020}, a moderate increase of voltage was already observed before a sharp jump even at low temperatures, in contrast to other samples. 
Thus, we concluded that the critical current density obtained for bp07 was far below the depairing limit and that it was dominated by the vortex intrusion into the nanobridge or some PS phenomena. 

Figure 4(a) shows the temperature dependence of $j_c$ for other 4 samples. 
Note that the values of $j_c$ for bp16 and bp18 at the lowest temperature, which are about a half of the value for ij12, are sufficiently larger than the in-plane depinning $j_c$. 
We found that the values of $T_c^J$ for both bp16 and bp18 were slightly higher than those of $T_c^R$, as shown in Table I. 
This is explained by the fact that $R_2(T)$ in both samples showed the multi-stage SC transition, which is considered to be induced by the slight spatial distribution of P dopants in the FIB microfabrication. This leads to a depressed value of $T_c^R$, since a rising temperature of $R_2(T)$ measurements is sensitive to a small part of the nanobridge with the lowest $T_c$. 
On the other hand, the value of $T_c^J$ is determined by $j_c$ at a threshold voltage with a sharp transition in the $I$-$V$ characteristics. The sharp transition is rather affected by a large part of the nanobridge showing the largest decrease in the $R_2(T)$ measurements. 
Thus, we used the reduced temperature normalized by $T_c^J$ in Fig.~4(b). 

In spite of the large difference of $T_c^J$ and $j_c^{\rm GL}(0)$ between bp06, bp16 and ij12, 
the plots of $j_c(T)/j_c(0)^{\rm GL}$ versus $T/T_c^J$ for these samples seem to fall into the almost identical curve in a broad temperature range from $T_c$ to $\sim$0.15~$T_c$, as shown in Fig.~4(b). 
This strongly suggests that the depairing $j_c$ along the $c$ axis can be directly determined down to 0.15~$T_c$ for these samples. 
Meanwhile, the plots for bp18 show a slightly depressed growth below 0.7~$T_c$. 
As shown in Table I, the lateral sizes ($w$ and $h$) and the length ($l$) of the $c$-axis nanobridge for bp18 is not so different from other samples, suggesting that the depressed $j_c$ is not derived from the nonuniform distribution of the supercurrent in the cross-sectional area. 
Thus, $j_c(T)$ for bp18 may be dominated by the PS events below 0.7~$T_c$, similar to the first jump for ij12 at low temperatures. 
Although we could not observe any step-like structure in the $I$-$V$ measurements for this sample, we consider that the oscillatory PS solution giving the step structure is not always stable in the condition of our measurements. 

In the previous studies on the measurements of the depairing current density \cite{Sun2020,Nawaz2013,Rusanov2004}, the smaller values of $(j_c/j_c^{\rm GL})^{2/3}$ than the theoretical values ($=0.53$ at $T$=0) predicted by the KL theory have often been discussed. Nawaz {\it et al.} \cite{Nawaz2013} and Rusanov {\it et al.} \cite{Rusanov2004} have focused on the current crowding effects around the corners of YBa$_2$Cu$_3$O$_y$ microbridges and the sample heating effects via contacts on Nb microbridges, respectively. 
In the $c$-axis nanobridges used in this study, there is a 90$^\circ$ turn near the current-gate, as shown in the lower inset of Fig.~1(b), which is expected to cause the current crowding \cite{Clem2011}. 
However, such an effect of the current crowding seems to be nearly common to all the measured samples, since the shape and size of the slits giving the 90$^\circ$ turn are all similar. 
Furthermore, the contacts of current lead in this study are far enough from the $c$-axis nanobridges, as shown in the upper inset of Fig.~1(b). 
Thus, both the current crowding and the sample heating via contacts never explain the depressed behavior of $j_c$ for bp18. 

On the other hand, Sun {\it et al.} have dealt with the multiband effects of Fe$_{1+y}$Te$_{1-x}$Se$_x$ \cite{Sun2020}. 
Such effects attract much attention as a possible explanation for a $T$ linear behavior observed in the temperature dependence of the upper critical field $\mu_0H_{c2}$ for magnetic field parallel to the $c$ axis \cite{Pan2024}.  
Although BaFe$_2$(As$_{1-x}$P$_x$)$_2$ is also a multiband system, the obtained results of $j_c$ along the $c$ axis showed a much closer behavior to the calculated curve by the KL theory than the previous results on Fe$_{1+y}$Te$_{1-x}$Se$_x$. 
Thus, the pair-breaking effects in nonequilibrium superconductivity including multiband effects should be microscopically understood to resolve the difference between BaFe$_2$(As$_{1-x}$P$_x$)$_2$ and Fe$_{1+y}$Te$_{1-x}$Se$_x$. 

The recent theoretical study for a dirty superconductor shows that the broadening of quasiparticle density of states (DOS), given by $\Gamma/\Delta_0$, reduces the depairing current density \cite{Kubo2020}. 
Here, $\Gamma$ is a Dynes parameter representing the broadening in the DOS peak. 
The comparison between such a calculation and our results suggests that $0.05 < \Gamma/\Delta_0 < 0.2$. 
It should be noted that the theoretical decrease of the depairing current density due to the finite value of $\Gamma/\Delta_0$ seems to appear in a whole temperature region below $T_c$, while our results suggest that the deviation from the KL theory becomes prominent below 0.7~$T_c$, as shown in Fig.~4(b).   
Therefore, we conclude that the most plausible candidate for explaining the observed difference from the KL theory is the influence of the PS phenomena. 
This also implies that the suppression of the PS oscillation giving the multi-staged structure in the $I$-$V$ curves is quite important for the precise determination of the depairing current density. 
Even for ij12 with the largest $j_c(0)^{\rm GL}$, it is possible that the second jump as well as the first jump are caused by the occurrence of PSCs. 
In this sense, it is preferable that the depairing current density is measured in the hysteretic region, since the zero-voltage state is expected to be broken at the depairing $j_c$ when we start from the SC state with $j<j_c$ in the hysteretic region \cite{Baranov2013}.

Finally, we refer to an interesting issue relevant to this work. 
As well known, BaFe$_2$(As$_{1-x}$P$_x$)$_2$ is considered to be one of ideal systems to investigate the signature of a quantum critical point (QCP) in the phase diagram \cite{Shibauchi2014}. 
The isovalent substitution of P for As does not induce serious disorders in the charge transport, as proved by the observation of clear quantum oscillations in a wide range of $x$ \cite{Shishido2010}.
To date, the experimental evidence of QCP located at $x_c\sim 0.3$ is given by several results including the non-Fermi liquid behaviors of the resistivity \cite{Kasahara2010}, the NMR relaxation rate \cite{Nakai2010}, and the anomalous enhancement of the effective mass, $m^*$ \cite{Shishido2010,Walmsley2013,Hashimoto2012}. 
Particularly, the $x$ dependence of London penetration depth ($\lambda_L\propto\sqrt{m^*}$) attracted much attention as the QCP beneath the superconducting (SC) dome \cite{Shibauchi2014,Hashimoto2012}. 
Recently, a similar peak was also observed in the electron-doped Ba(Fe$_{1-x}$Co$_x$)$_2$As$_2$. This indicates that the signature of QCP is not limited in the clean sytem, since Co-doped BaFe$_2$As$_2$ is considered to be significantly disordered by electron doping, showing a sharp contrast to P-doped BaFe$_2$As$_2$ \cite{Joshi2020}. 
We note that the depairing current density given by the GL theory is dependent on $m^*$, since $j_c(\propto\xi\lambda_L^2)\propto 1/\sqrt{m^*}$, while a critical velocity depairing Cooper pairs is proportional to $\Delta/k_F$ in the BCS theory, where $\Delta $ is the superconducting gap and $k_F$ is a Fermi wavenumber, respectively. 
Thus, it is expected that the $x$ dependence of the depairing current density in both  
BaFe$_2$(As$_{1-x}$P$_x$)$_2$ and Ba(Fe$_{1-x}$Co$_x$)$_2$As$_2$ will provide an important insight to settle this issue.

\section{Conclusion}
In conclusion, we investigated the $c$-axis charge transport of BaFe$_2$(As$_{1-x}$P$_x$)$_2$ single crystals by using the $c$-axis nanobridge structure, additionally fabricated into the narrow $ab$-plane bridge. 
This technique is excellent for obtaining the interlayer resistivity of soft materials such as BaFe$_2$(As$_{1-x}$P$_x$)$_2$, where small cracks are partially introduced by cutting and exfoliating crystals. 
Although the additional FIB etching brought the slight decrease of P concentration in the narrow bridges of BaFe$_2$(As$_{1-x}$P$_x$)$_2$, the normal-state resistivity anisotropy in the slightly underdoped side near the optimal doping was found to be at most 8 just above $T_c$, showing a good agreement with the previous results for bulk crystals. 

We also confirmed that the magnitude of $j_c$ along the $c$ axis determined by the $I$-$V$ characteristics was much larger than the $ab$-plane depinning $j_c$ determined by the previous magnetization measurements. 
The temperature dependence of $j_c$ near the optimal doping showed a good agreement with the KL theory, indicating the achievement of the depairing limit along the $c$ axis. 
These results ensure that we can explore the $x$ dependence of the depairing $j_c$ of BaFe$_2$(As$_{1-x}$P$_x$)$_2$ and Ba(Fe$_{1-x}$Co$_x$)$_2$As$_2$ by using the $c$-axis nanobridge, in order to settle the unresolved issue of quantum critical behavior.  

In addition, we observed the two-staged voltage jumps in the $I$-$V$ characteristics at low temperatures, strongly suggesting the contribution of phase-slip phenomena near the depairing limit. 
The slight deviation of the depairing $j_c$ from the KL theory observed at low temperatures may also be attributed to the contribution of phase-slip phenomena. 
Although such phenomena are still far from the complete understanding, 
our results manifestly show that they occur not only in narrow wires and two-dimensional strips but also in three-dimensional bridges with small cross-sectional area. 
Detailed studies on the $I$-$V$ characteristics near the depairing limit will provide useful information to approach the complete comprehension.

\acknowledgments
This work was partly supported by KAKENHI from JSPS (JP20H05164) and Aoyama Gakuin University Research Institute grant programs for research unit and creation of innovative research. 
FIB microfabrication in this work was supported by Center for Instrumental Analysis, College of Science and Engineering, Aoyama Gakuin University.

\bibliography{references_ba122opt_final}

 \begin{figure}
 \includegraphics[width=16cm]{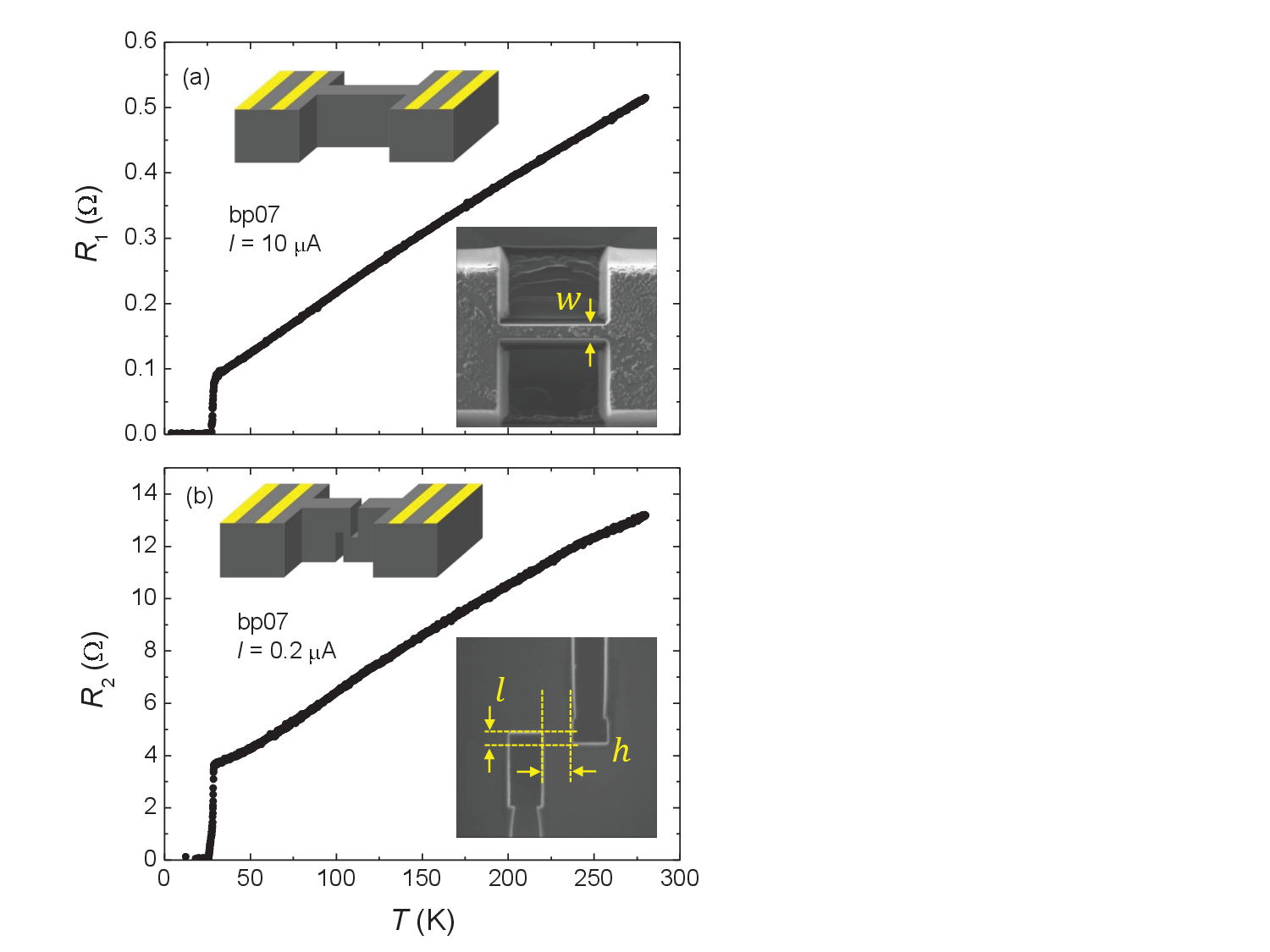}%
 \caption{(a) Temperature dependence of the resistance ($R_1$) for the microbridge in the $ab$-plane. Upper and lower insets are a schematic viewgraph and a scanning ion microscope (SIM) image of the microbridge, respectively. Here, $w$ is a width of the microbridge. (b) Temperature dependence of the resistance ($R_2$) for the nanobridge with two additional slits along the $c$ axis. Upper and lower insets are a schematic viewgraph and a SIM image, respectively. Here, $h$ and $l$ are the distance between the two slits and the length of the $c$-axis nanobridge, respectively. }\label{fig1}
\end{figure}

 \begin{figure}
 \includegraphics[width=20cm]{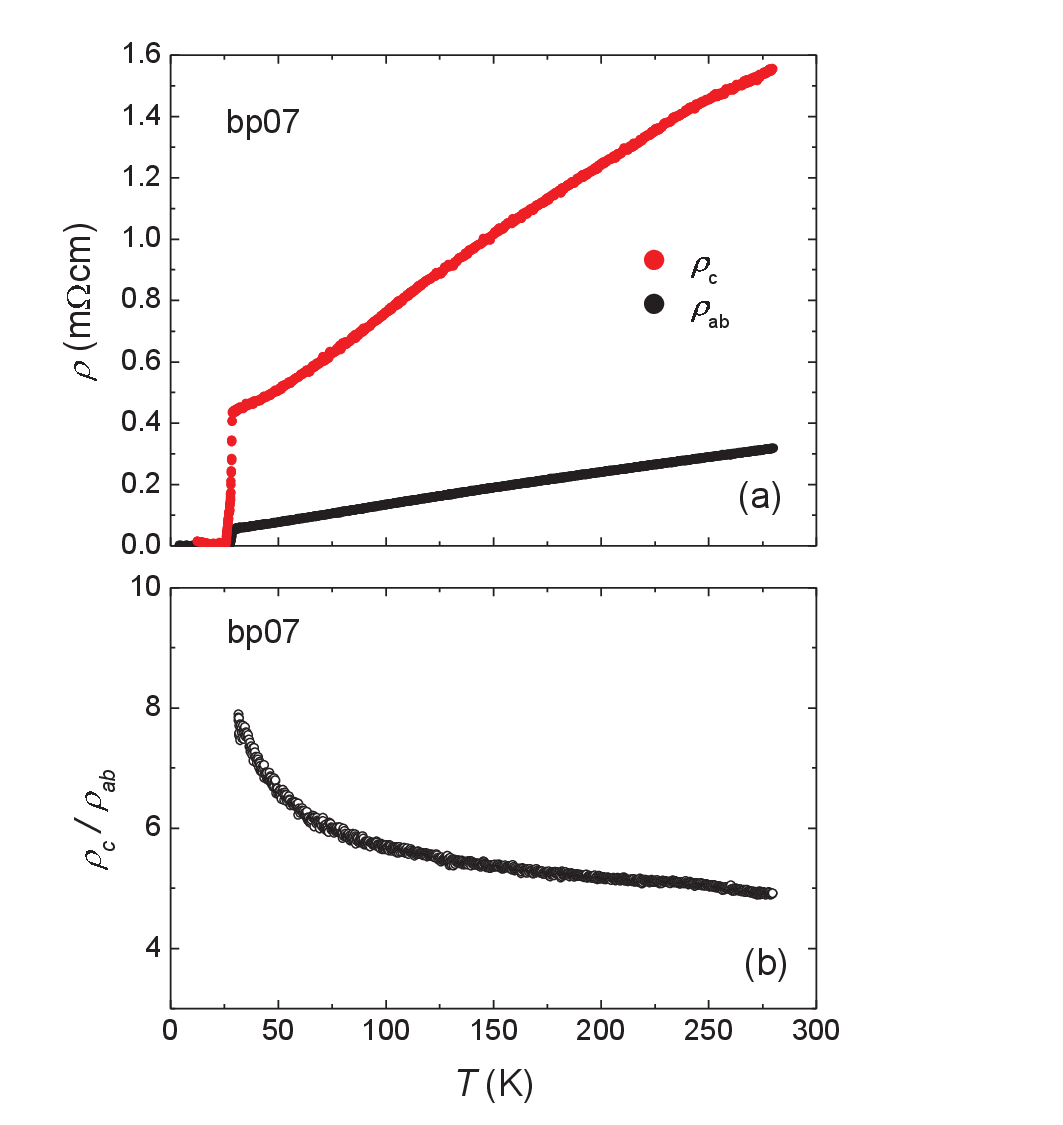}%
 \caption{(a) Temperature dependence of the resistivity in the $ab$-plane (black circles) and along the $c$ axis (red circles), which were obtained from $R_1$ and $R_2$. (b) Temperature dependence of the normal-state resistivity anisotropy. }\label{fig2}
\end{figure}

 \begin{figure}
 \center
 \includegraphics[width=18cm]{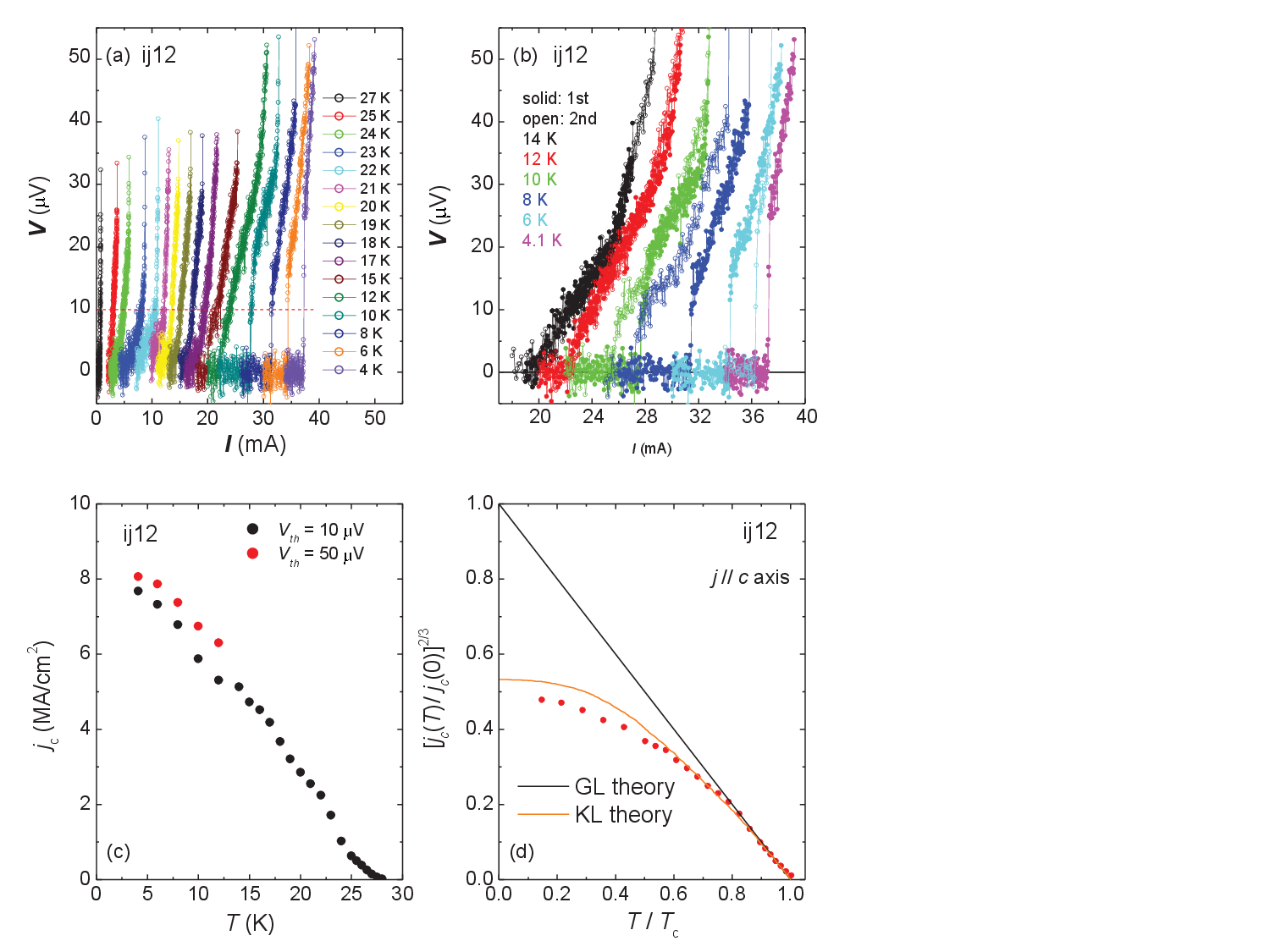}%
 \caption{(a) The current-voltage ($I$-$V$) characteristics for the $c$-axis nanobridge (ij12), measured at several temperatures below $T_c$. (b) The $I$-$V$ curves below 14~K. Solid symbols are the same data as those in (a), while open symbols represent the 2nd $I$-$V$ measurements. (c) Temperature dependence of the $c$-axis critical current density, which is determined from the 1st measurements by using a threshold voltage at 10~$\mu$V (black circles) or 50~$\mu$V (red circles). (d) Reduced temperature dependence of the $c$-axis critical current density, normalized by a value of $j_c^{\rm GL}(0)$. Black and orange lines are the calculation of the GL and KL theories, respectively. }\label{fig3}
\end{figure}

 \begin{figure}
 \center
 \includegraphics[width=18cm]{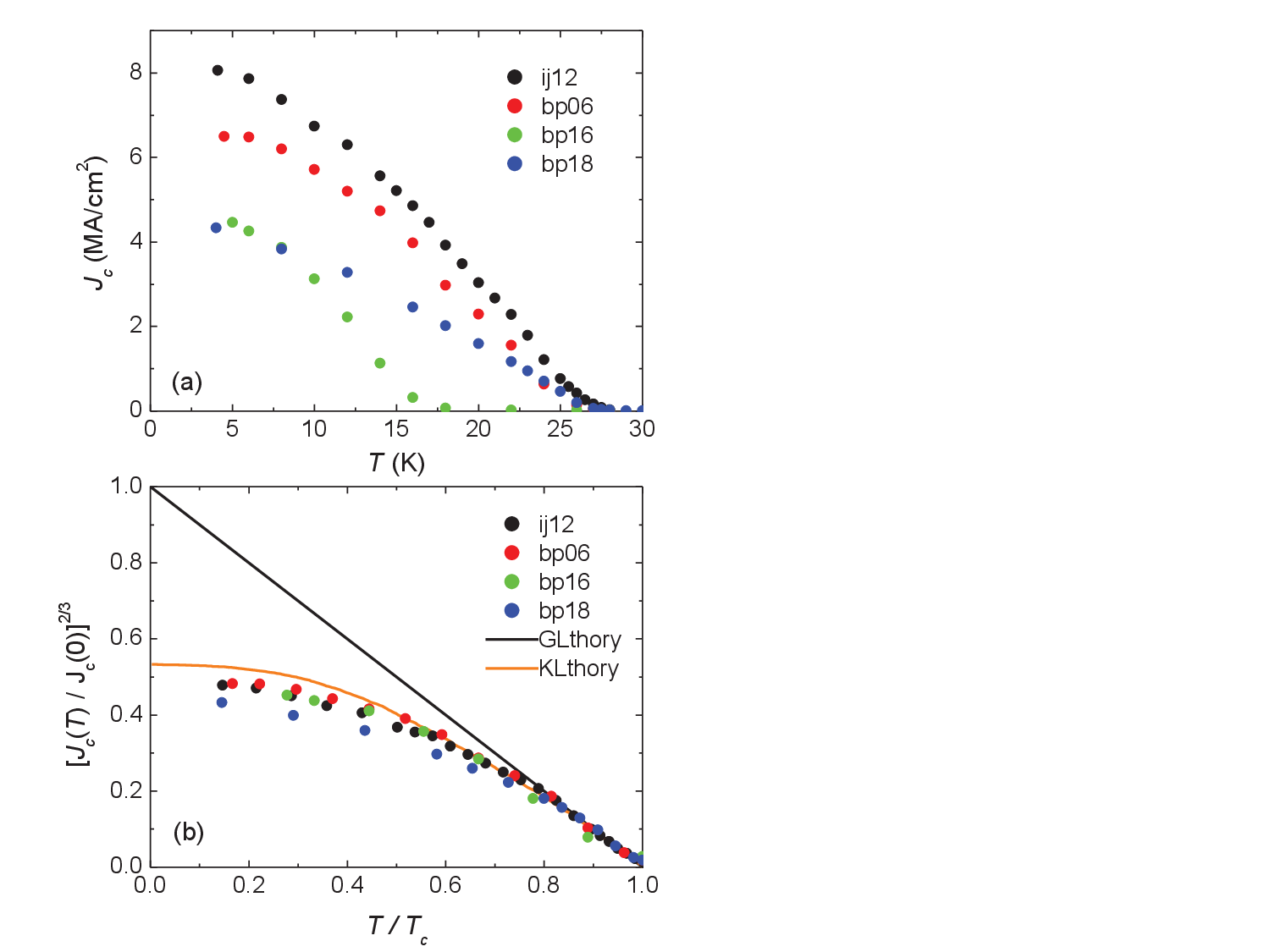}%
 \caption{(a) Temperature dependence of the $c$-axis critical current density for all samples. (b) Reduced temperature dependence of the $c$-axis critical current density, normalized by a value of $j_c^{\rm GL}(0)$. The reduced temperature is obtained by using $T_c^J$. Black and orange lines are the calculation of the GL and KL theories, respectively. }\label{fig4}
\end{figure}


\end{document}